\documentstyle[amssymb,latexsym,psfig]{ag}

\title[Star formation time-scales in M82]{Star formation time-scales in
the nearby, prototype starburst galaxy M82}

\author[R.  de Grijs]{Richard de Grijs\\
Institute of Astronomy, University of Cambridge, Madingley Road,
Cambridge CB3 0HA; E-mail: grijs@ast.cam.ac.uk}

\pubyear{2001}

\begin{document}
\maketitle

\begin{abstract}
The last tidal encounter between M82 and M81, some 500 Myr ago, had a
major impact on what was probably an otherwise normal, quiescent disc
galaxy.  It caused a concentrated burst of star formation in the form of
massive star clusters, which decreased rapidly, within a few 100 Myr.
The current starburst in the centre of the galaxy is likely either due
to large-scale propagating star formation or possibly related to late
infall of tidally disrupted debris from M82 itself.  It may, in fact, be
a combination of these two mechanisms, in the sense that the star
formation in the active core is actually propagating, while the overall
evolution of the starburst is due to tidal debris raining back onto the
disc of the galaxy, causing the present-day starburst.
\vspace*{0.5cm}
\end{abstract}


\section{M82, the prototype starburst galaxy}

M82 is often regarded as the ``prototype'' starburst galaxy, being the
nearest and best-studied example of this class of galaxies. 
Observations over the entire wavelength range, from radio waves to
X-rays (reviewed in Telesco 1988, Rieke et al.  1993), seem indicative
of the following scenario.  During the last several 100 Myr, tidal
interactions with M81, and perhaps also with other galaxies in the same
group, induced intense star formation in M82 due to increased gas flows
channeled into its centre.  The resulting starburst, with the high star
formation rate of $\sim 10 M_\odot$ yr$^{-1}$, has continued for up to
about 50 Myr.  Energy and gas ejection from supernovae and combined
stellar winds drive a large-scale galactic wind along the minor axis of
M82 (e.g., Lynds \& Sandage 1963, McCarthy et al.  1987, Shopbell \&
Bland-Hawthorn 1998). 

All of the bright radio and infrared sources associated with the active
starburst are confined in a small region within a radius of $\sim 250$
pc from the galaxy's centre; the highest surface brightness regions in
the active core, M82 A and C, are indicated in Fig.  1.  Most of this
volume is heavily obscured by dust at optical wavelengths. 

\begin{center}
\begin{figure*}
\psfig{figure=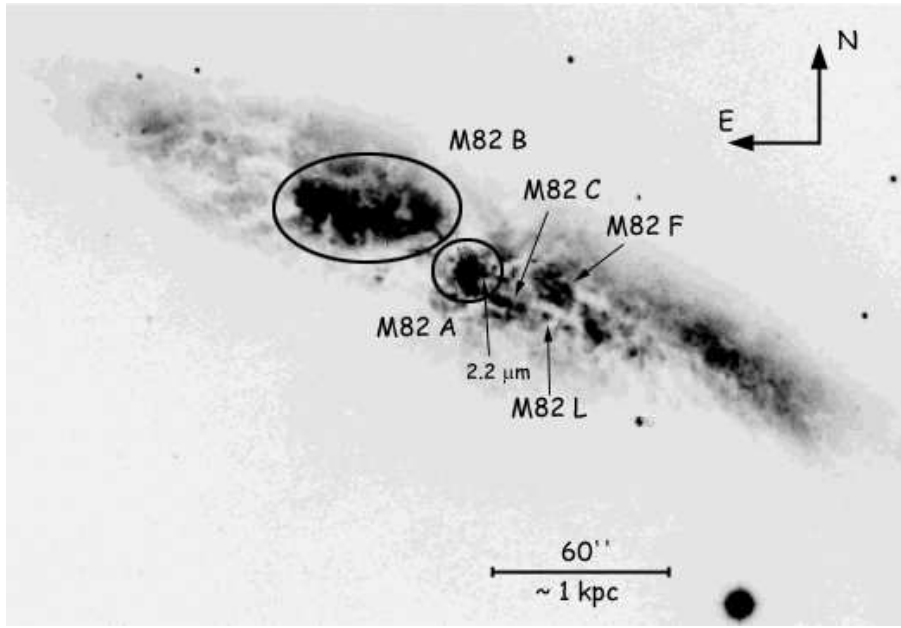,width=12cm}
\caption[]{M82 {\it B}-band image obtained at the Palomar 5m telescope,
taken by A.  Sandage (exposure time 20 minutes, seeing $\lesssim 1''$). 
The locations of the active starburst core in M82 A and C, and the older
starburst region M82 B are indicated, as well as star clusters F and L. 
We have also indicated the location of the 2.2 $\mu$m peak coincident
with the galactic centre; the 390 pc molecular spiral arms studied by
Shen \& Lo (1995) are located at $15-20''$ on either side of the
2.2$\mu$m peak along the galaxy's major axis.}
\end{figure*}
\end{center}

However, there is now ample evidence that the active starburst was not
the only major starburst event to have occurred in M82.  A region at
$\sim 400-1000$ pc from the centre, M82 B (Fig.  1), has the high
surface brightness and spectral features expected for a fossil starburst
with an age $\gtrsim 100$ Myr and an amplitude similar to the active
burst (O'Connell \& Mangano 1978, Marcum \& O'Connell 1996, de Grijs et
al.  2001).  Its A-star dominated spectrum is characterised by Balmer
absorption lines, and exhibits a large Balmer discontinuity.  Emission
lines in M82 B are very weak, as opposed to the active starburst, which
features intense line emission, in particular of the Balmer lines
(Marcum \& O'Connell 1996).  These are the defining characteristics of
the anomalous ``E+A'' spectra found in distant galaxy clusters, which
are generally interpreted as the signature of a truncated burst of star
formation that occurred $100-1000$ Myr earlier (e.g., Couch \& Sharples
1987, Dressler \& Gunn 1990, Couch et al.  1998).  These starbursts are
probably part of the process by which disc galaxies are converted into
elliptical or lenticular galaxies and are thought to result from tidal
interactions, mergers, or perhaps ram-pressure stripping by the
intergalactic medium (Butcher \& Oemler 1978, Oemler 1992, Barger et al. 
1996). 

The significance of the detailed study of M82's starburst environment
lies, therefore, in the broader context of galaxy evolution.  Starbursts
of this scale are likely to be common features of early galaxy
evolution, and M82 is the nearest analogue to the intriguing sample of
star-forming galaxies recently identified (e.g., the so-called Lyman
break galaxies) at high redshifts ($z \gtrsim 3$; Steidel et al.  1996,
Giavalisco et al.  1997, Lowenthal et al.  1997).  M82 affords a
close-up view not only of an active starburst but also, in region B and
elsewhere, of the multiple post-burst phases.  Other nearby galaxies are
known to exhibit one or another of these features, but none of these
offers the opportunity to study both at such close range, with a
comparably high spatial resolution, or with such a wealth of correlative
data, as M82. 

\section{Tidally-induced starburst activity -- do the time-scales work
out?}

Although the M82 starburst regions are pervaded by high-extinction
filaments, the outermost parts of the starburst core have lower
extinction and can be studied with optical telescopes.  However, the use
of optical and even near-infrared wavelengths effectively limits our
sampling of the M82 starbursts to their surface regions.  With this
caveat in mind, one can derive the approximate age distribution of the
major starburst events in M82 using a variety of tracer objects and
methods. 

\subsection{The Mean Stellar Population in the Disk of M82}

Rieke et al.  (1993) argued, based on the modelling of integrated,
ground-based spectra of the central regions of M82, that the galactic
centre has most likely been dominated by two starburst events in the
past $\sim 30$ Myr, spaced by about 25 Myr, each of which had a duration
of $\sim 5$ Myr.  Detailed CO observations and Br$\gamma$ equivalent
width measurements (Satyapal et al.  1997) confirm the mean age of the
active starburst to be roughly 10 Myr, although this depends to some
degree on the detailed time dependence of the star formation rate. 

From 20--40 {\AA} resolution ground-based spectrophotometry, Marcum \&
O'Connell (1996) showed that the spectrum of the active starburst is in
fact dominated by a very young, $\sim 5$ Myr old stellar population,
superimposed on an older population with a late-A/early-F star main
sequence turn-off and a red-giant clump, corresponding to an age of
$\sim 600$ Myr. 

Away from the active centre, in M82 B, the earliest-type stars have
evolved off the main sequence, leaving a dominant stellar population
with a late-B/early-A star main sequence turn-off, corresponding to an
age of $\sim 100$ Myr (O'Connell \& Mangano 1978, Marcum \& O'Connell
1996), with possibly an older underlying stellar population. 

\begin{figure*}
\psfig{figure=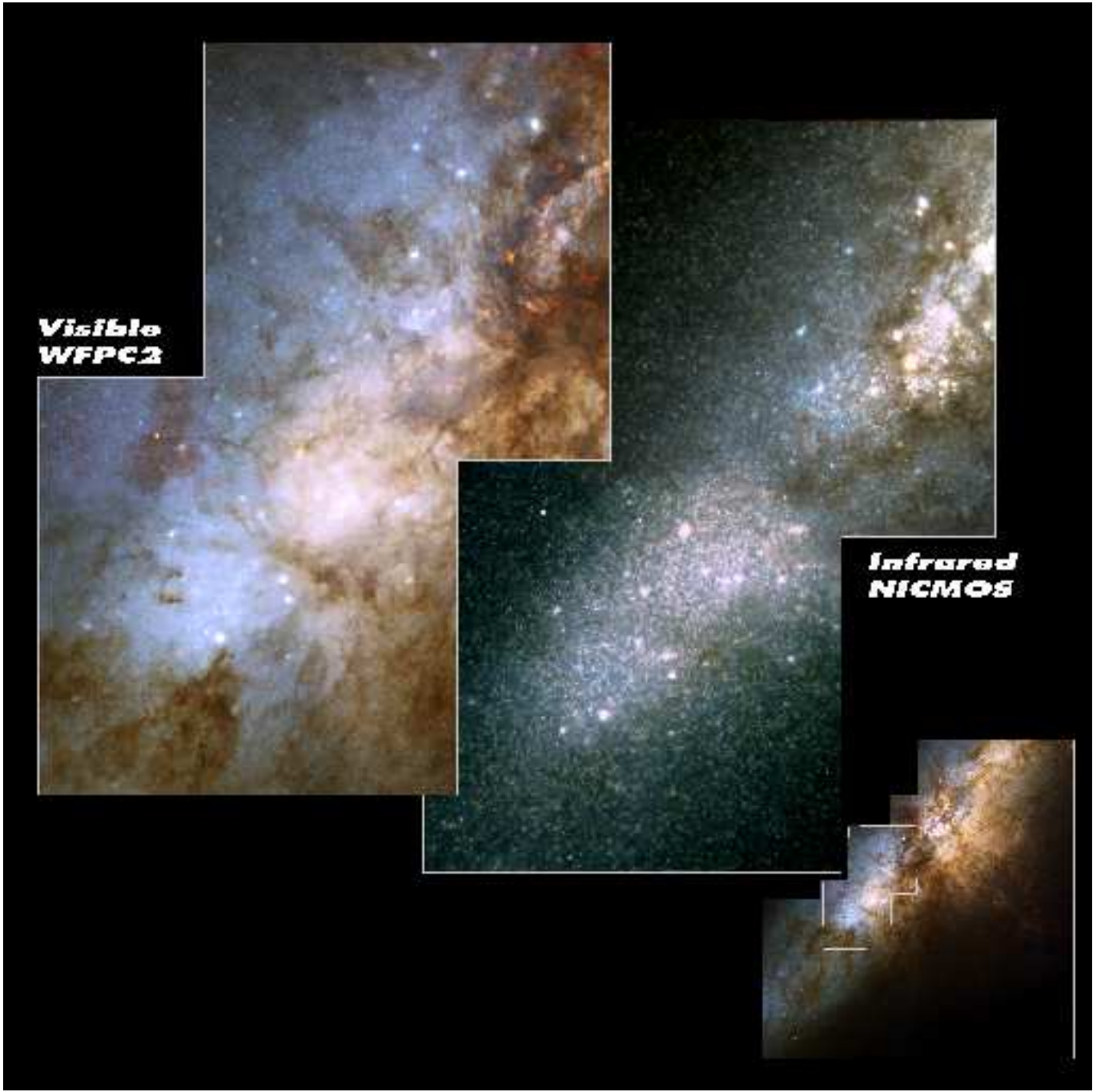}
\caption[]{False-colour images of M82 obtained with the {\sl Hubble
Space Telescope} in the optical ({\sl WFPC2}) and near-infrared
(NICMOS).  It is immediately clear that at optical wavelengths the
effects of the ubiquitous dust are severe, thus restricting optical
studies to the region's surface area.  In the near-infrared, large
numbers of red giant stars appear, thus allowing us -- for the first
time -- to resolve the disc stellar population in M82.  Inset: Full {\sl
WFPC2} field of view centred on M82 B.  Source: ESA press release, 7
March 2001.}
\end{figure*}

Using the superb resolution of the Hubble Space Telescope, we were able
to resolve the galactic background of M82 B into individual stars for
the first time: it is predominantly composed of faint point sources, and
these become increasingly dominant at longer (near-infrared)
wavelengths, as is clearly illustrated in Fig.  2.  Although our
magnitude cut-off prevents us from detecting the upper main sequence or
giants older than about 80 Myr, a comparison with solar-metallicity
Padova isochrones indicates that the brightest M82 giant stars are core
helium-burning stars with ages $\sim 20-30$ Myr (de Grijs et al.  2001;
Fig.  3).  Significant star formation has not occurred in M82 B in the
last 10--15 Myr.  We based this on a comparison with late-type, young
($\sim 10$ Myr) K and M supergiants in the Large Magellanic Cloud, which
are clearly offset from the bulk of the M82 disc stars.  Only 4.3\% of
the integrated near-infrared light originates in the resolved
population, implying that the great majority of cool stars were formed
at earlier times.  For completeness, we mention that the age estimates
would not change significantly for metal abundances of $0.4-1.2 \times$
solar, the likely range for M82. 

\begin{figure}
\psfig{figure=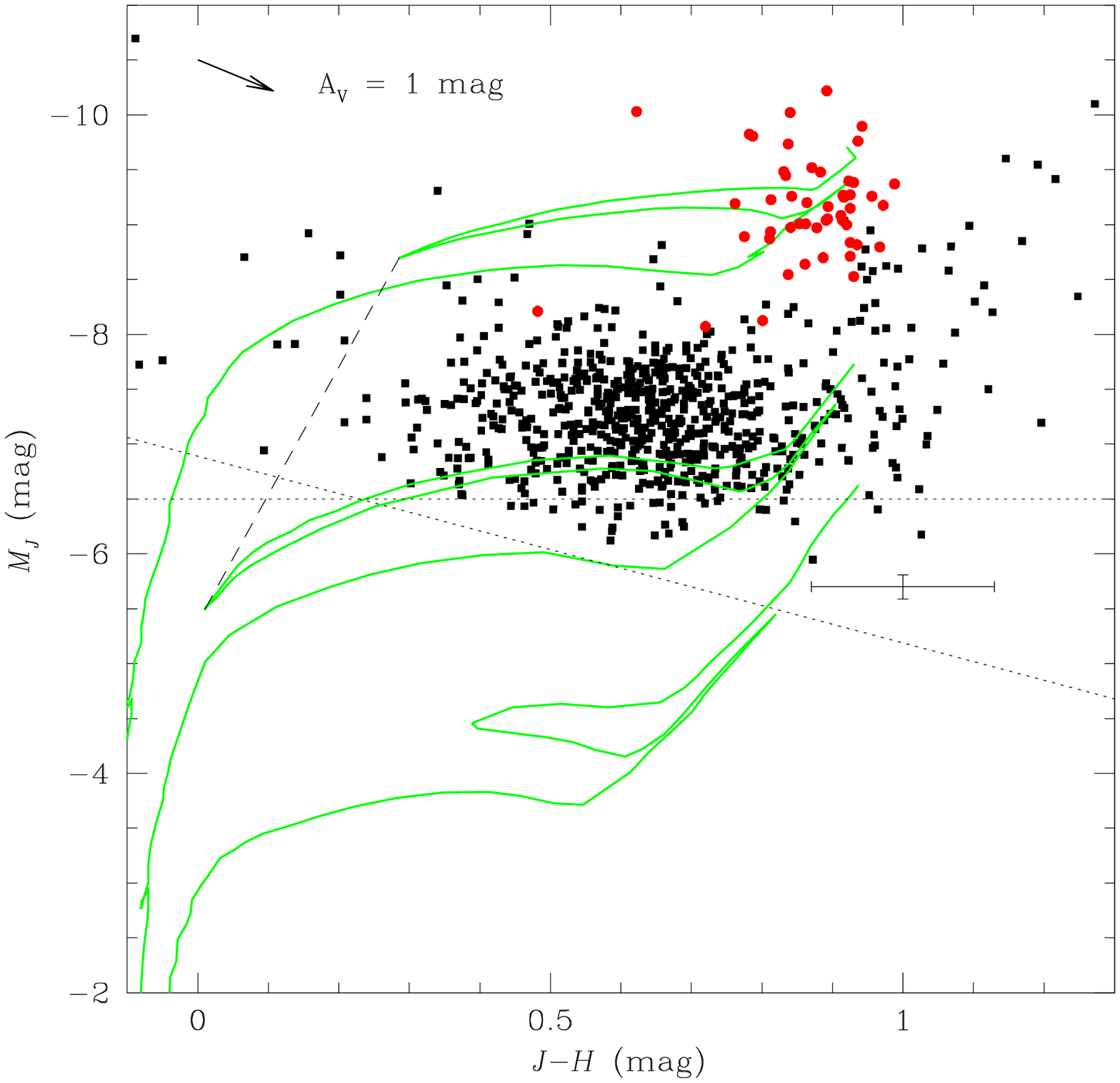,width=9cm}
\caption[]{Near-infrared colour-magnitude diagram for the stellar
background in the disc of M82.  The green lines represent theoretical
Padova isochrones for ages of 10, 30, and 100 Myr (top to bottom) and
solar metallicity.  The dotted lines are the observational selection
limits; the dashed lines indicate the envelope of the regions occupied
by the theoretical isochrones between 10 and 30 Myr.  The scatter in the
data points towards the red of the reddest colours reached by the
isochrones is most likely due to the effects of dust, as inferred from
the direction of the extinction vector.  For the same reason, the area
close to the blue envelope (dashed line) is underpopulated.  Typical
observational uncertainties are indicated by the error cross.  We have
also included K and M supergiants in the Large Magellanic Cloud (red
data points).  Obviously, field star formation has continued in M82 B
until $\sim 20$ Myr ago.  It has evidently been suppressed during the
last $\sim 10-15$ Myr.}
\end{figure}

\subsection{Young Compact Star Clusters}

{\sl Hubble Space Telescope} {\it V} and {\it I}-band imaging of bright
optical features in the direction of the core revealed over 100 young
compact star clusters (sometimes referred to as ``super star clusters'')
with $M_V \sim -12$, brighter than any star cluster in the Local Group
(O'Connell et al.  1995).  Ages of these clusters are estimated to be
$\sim 10-50$ Myr (O'Connell \& Mangano 1978, O'Connell et al.  1995). 
Satyapal et al.  (1997) focused on the highest-luminosity objects in the
central starburst at near-infrared wavelengths, assuming that these
represent embedded young star clusters deep inside the starburst core. 
Their best age estimates for these objects, assuming a single formation
event, is about 10 Myr, in good agreement with optical age estimates of
star clusters and in general agreement with M82's central star formation
history. 

Two of the bright optical structures near the centre seen on
ground-based images deserve further discussion.  The brightest star
forming region, M82 A (cf.  O'Connell \& Mangano 1978), corresponds
spatially with the conspicuous nuclear peak observed in the
near-infrared, at 2.2 $\mu$m (see Fig.  1).  Ground-based imaging and
comparison with stellar population models yields a fairly robust age
estimate for the mean age of M82 A of $\lesssim 50$ Myr (O'Connell \&
Mangano 1978, O'Connell et al.  1995).  {\sl Hubble Space Telescope}
images resolve this star forming knot into more than 50 individual
luminous star clusters, superimposed on a bright unresolved background. 

Using high-resolution (1.6 {\AA}) ground-based spectroscopy, Gallagher
\& Smith (1999) recently obtained an age of $60 \pm 20$ Myr for the very
luminous ($M_V \simeq -16$) cluster F, located 440 pc from the galaxy's
centre, and a similar age for the nearby, highly reddened cluster M82 L. 
While M82 F is only a single star cluster, it is very massive ($\sim 1.2
\times 10^6 M_\odot$; Gallagher \& Smith 1999, Smith \& Gallagher 2001),
so its formation would likely have had a noticeable effect on a
significant fraction of this extraordinary galaxy: Gallagher \& Smith
(1999) derive a high local star formation rate, of $> 1 M_\odot$
yr$^{-1}$, to produce this cluster in a dynamical time of about 1 Myr. 
However, its high radial velocity with respect to the galaxy's system
velocity indicates a highly eccentric orbit (Smith \& Gallagher 2001),
which must have caused M82 F to travel far from its initial locus in the
galaxy.  Therefore, we cannot be sure where exactly in M82 this massive
star cluster was actually formed. 

\begin{figure}
\psfig{figure=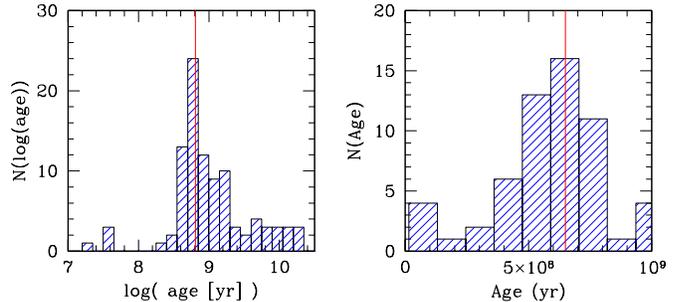,width=9cm}
\vspace{-4.5cm}
\caption[]{Distribution of compact star cluster ages in M82 B.  There is
a strong peak of cluster formation at $\sim 650$ Myr ago (red lines) but
very few clusters are younger than 300 Myr.  The left-hand panel shows
the age distribution of all star clusters on a logarithmic scale; the
right-hand panel is a zoomed-in version of the peak cluster formation
epoch.}
\end{figure}

Under the assumption that the older starburst region M82 B is analogous
to an evolved and therefore faded version of the active burst in the
core, we expected to find a similar, but slightly evolved population of
compact star clusters as found in the active starburst core.  Indeed, we
detected another $\sim 110$ young compact star clusters in M82 B with
intrinsic luminosities of $M_V \gtrsim -9$ (de Grijs et al.  2001), for
which we estimated ages from $\sim 30$ Myr to over 10 Gyr, with a peak
near 650 Myr (see Fig.  4), using broad-band optical and near-infrared
images (de Grijs et al.  2001).  About 22\% of the clusters in M82 B are
older than 2 Gyr, with a flat distribution to over 10 Gyr.  There is a
strong peak of cluster formation at $\sim 650$ Myr ago but very few
clusters are younger than 300 Myr (cf.  Fig.  4).  The full-width of the
peak is $\sim 500$ Myr, but this is surely broadened by the various
uncertainties entering the age-dating process.  The selection bias of
the star clusters in de Grijs et al.  (2001) is such that the truncation
of cluster formation for ages $< 200$ Myr is better established than the
constant formation rate at ages $> 2$ Gyr. 

Thus, we suggest steady, continuing cluster formation in M82 B at a very
modest rate at early times ($> 2$ Gyr ago) followed by a concentrated
formation episode lasting from 400--1000 Myr ago and a subsequent
suppression of cluster formation.  This leads us to conclude that M82 B
has evidently not been affected by the more recent ($< 30$ Myr)
starburst episode now occurring in the central regions. 

\subsection{Supernova Remnants}

The starburst core of M82 is well known to contain a large population of
compact supernova remnants (SNRs).  These are obscured at optical
wavelengths by large amounts of dust, but their structures and evolution
have been studied at radio wavelengths (e.g., Kronberg et al.  1985,
Huang et al.  1994, Muxlow et al.  1994, Golla et al.  1996, Allen \&
Kronberg 1998).  Based on their sizes and assuming a constant expansion
rate, their ages have been estimated at less than a few 100 yr (Huang et
al.  1994, Muxlow et al.  1994).  Although this does not set tighter
constraints on the age of the entire region than spectroscopic age
estimates, or age determinations derived from the young star cluster
populations, it confirms our view that the core is significantly younger
than regions at larger radii, which are largely devoid of young SNRs or
supernovae. 

However, even though the starburst core of M82 is pervaded by large
amounts of dust, thus making optical studies difficult, one may expect
to be able to look into the core itself along certain lines of sight,
where there are holes in the dust.  It is therefore not unlikely that
one would be able to detect the presence of SNRs based on their
characteristic ratio of [S II]/H$\alpha$ line emission. 

H$\alpha$ images, obtained with the {\sl Hubble Space Telescope} as well
as ground-based images (e.g., Lynds \& Sandage 1963, McCarthy et al. 
1987), show that there is relatively little line emission in the disc of
M82 at radii larger than 500 pc from the central starburst.  The easily
visible H$\alpha$ line emission is concentrated to the core starburst
region and the bright minor-axis superwind extending from it. 

As mentioned before, despite the high extinction in the starburst core,
O'Connell et al.  (1995), using optical imaging, found more than 100
young compact star clusters in this area.  Since these clusters are most
likely of order 10--50 Myr old, it is expected that many heavy stars
will already have ended in a supernova explosion.  Although some of the
roughly 50 radio supernovae and SNRs in M82 (Muxlow et al.  1994, Huang
et al.  1994) are expected to be associated with the young star
clusters, none appears to coincide with the star clusters seen with the
{\sl Hubble Space Telescope} (Golla et al.  1996).  A simple,
back-of-the-envelope calculation shows that in a population of 100 star
clusters of ages similar to those estimated in M82 A and containing
between $10^5$ and $10^6$ stars, one would expect to detect between
about 5 and $\sim 50$ type II SNRs at any given moment, assuming any
reasonable range of initial mass functions.  The question remains,
therefore, why none of the optically-detected young compact star
clusters show any evidence for the presence of SNRs. 

Golla et al.  (1996) suggested that the visible young star clusters in
M82 are located in the foreground and outside appreciable concentrations
of interstellar gas so that the supernova explosions were unable to
sweep up gas and form SNRs.  However, this does not explain why we do
not see any SNRs associated with the visible star clusters.  It would
also imply that 1500--3000 young star clusters exist in the core of the
galaxy, which seems an unrealistically high number.  Therefore, they
suggest that the detected radio supernovae and SNRs are hidden behind
dense layers of dust so that the associated star clusters are not seen. 

This argument has apparently gained support from recent observations of
HI absorption towards many radio supernovae and SNRs in M82 (Wills et
al.  1998), the amount of which seemed to imply that all radio-detected
sources in M82 are hidden behind an impenetrable layer of dust
(estimated mean extinction $\langle A_V \rangle = 24 \pm 9 mag$; Mattila
\& Meikle 2001).  However, this assumes that all objects detected at
radio wavelengths are associated with, or are hidden behind, dense
molecular clouds, which does not seem plausible: O'Connell et al. 
(1995) point out that the geometrical arrangement of the young compact
star clusters resembles the structure of the starburst so that at least
some of the observed clusters are located far inside M82.  In fact, they
derived a mean extinction towards M82 A of about 3 magnitudes, while
Satyapal et al.  (1997) argue that the 12 most luminous compact
near-infrared sources are deeply embedded star clusters in the starburst
core, thus also implying optical depths $< 1$ at near-infrared ({\it K}
band) wavelengths along these sight lines.  In addition, the extinction
measured toward M82 F is only about 5 magnitudes although the cluster is
located far inside M82 (Gallagher \& Smith 1999). 

Therefore, we suggest instead that the conditions in and near young star
clusters may be particularly hostile at least for the formation of SNRs
(Greve et al.  2001), i.e., the matter ejected by supernovae is quickly
dispersed because of stellar winds, nearby supernova explosions, and the
strong gravitational field of the star clusters.  In addition, the
interstellar medium in these clusters is possibly dense (Huang et al
1994), thus limiting SNRs to very small sizes (Huang et al.  1994,
Muxlow et al.  1994).  It is therefore possible that in and near young
star clusters there are only a few short-lived radio supernovae but no,
or very few, and small, SNRs. 

Although we do not expect to detect SNRs in the older starburst region,
M82 B, simply because of its estimated mean age of a few 100 Myr, we do
detect diffuse H$\alpha$ emission, but at much lower levels than in the
active starburst and in the bright superwind (de Grijs et al.  2000). 

Clearly, many of the optically detected compact star clusters (de Grijs
et al.  2001) have little or no H$\alpha$ emission, which is not
unexpected if these clusters are indeed older than a few 100 Myr. 
However, some compact H$\alpha$ emission regions have inconspicuous
continuum counterparts.  In addition, there seems to be a gradient in
H$\alpha$ emission within M82 B: there are few compact H$\alpha$ sources
far away from the central starburst, whereas both compact sources and
diffuse emission are seen nearer to the starburst core, although still
at a surface brightness $20 \times$ lower than in the active starburst.

One expects to find two types of compact H$\alpha$ sources in a galaxy
like M82.  H{\sc ii} regions will exist around young star clusters with
ages $\lesssim 10$ Myr because of the presence of ionizing O- or early
B-type stars.  Even for very young H{\sc ii} regions, including those
that suffer from significant extinction, there should always be a
significant continuum source associated with these.  Alternatively, Type
II supernovae can also produce compact H$\alpha$ remnants.  These will
often be associated with their parent star clusters, but in many cases
there may be no well-defined compact continuum source.  Since Type II
supernovae can occur up to 50 Myr after a star formation event, the
associated cluster may have faded due to evolutionary effects or
dynamically expanded enough to be inconspicuous against the bright
background of the galaxy.  Alternatively, the parent star of the SNR
could have been ejected from the cluster or could have formed initially
in the lower density field stellar population.  In fact, studies of
resolved starbursts suggest that 80--90\% of the bright stellar
population resides in a diffuse component outside of compact clusters
(e.g., Meurer et al.  1995, O'Connell et al.  1995), which is partially
replenished by the dissolution of star clusters on time-scales of $\sim
10$ Myr (cf.  Tremonti et al.  2001). 

We presented acceptable evidence that the 10 most luminous compact
H$\alpha$ sources in M82 B could be SNRs (de Grijs et al.  2000).  Six
of these have only faint counterparts in the continuum passbands and
some of the sources show evidence of limb brightening, as might be
expected for older SNRs.  Their H$\alpha$ luminosities, surface
brightnesses and sizes are consistent with those of SNRs in other
star-forming galaxies.  However, none of these H$\alpha$-bright SNR
candidates, with one possible exception, appear to show 8.4 GHz radio
emission brighter than our detection limit associated with them. 
However, most of the bright 8.4 GHz sources are located in areas with
higher than average extinction, where detection of optical continuum or
H$\alpha$ counterparts would be difficult. 

It is not unusual for SNRs to be H$\alpha$ bright while lacking
significant emission at radio wavelengths.  Radio and H$\alpha$ emission
are due to different physical processes and are prominent at different
evolutionary stages of the SNR; radio emission arises earlier in SNR
evolution than H$\alpha$ emission. 

The presence of SNRs in M82 B can help to set limits on its star
formation history.  The last supernovae in a quenched starburst would
occur at a time comparable to the longest lifetime of a supernova
progenitor after the end of the starburst activity.  The time spent
between the zero-age main sequence and planetary nebula phase by an $8
M_\odot$ progenitor star, which is generally adopted as a lower limit
for Type II supernovae (e.g., Kennicutt 1984), corresponds to $\sim
35-55$ Myr (Iben \& Laughlin 1989, Hansen \& Kawaler 1994).  Type Ia
supernovae, which involve lower mass stars in binary systems, can occur
much later, but one expects these to be more uniformly distributed over
the galaxy's surface, not concentrated near regions of recent star
formation. 

Therefore, if our candidates are indeed SNRs they suggest an upper limit
to the end of the starburst event in M82 B closest to the active
starburst core of $\sim 50$ Myr.  The evidence that these objects are
actually SNRs is only circumstantial, however.  Although their
properties seem to compare better to SNR samples in other galaxies than
to those of H{\sc ii} regions, they could be H{\sc ii} regions which
have been affected by the unusual physical circumstances in M82's disc. 
A straightforward test would be to obtain [S II]/H$\alpha$ emission line
ratios for these sources, since these are sensitive to the presence of
strong shock waves. 

\section{The overall star formation history of M82}

These results suggest that starburst activity in M82 is of longer
duration than had been supposed and has possibly propagated through the
galaxy's disc (Shen \& Lo 1995, Satyapal et al.  1997, de Grijs et al. 
2000).  Combining the wealth of observational information for the M82
starbursts, the following picture for its star formation history
emerges. 

The observed distribution of gas in the M81/M82 group of galaxies is
consistent with a 3-body model, including M81, M82 and NGC 3077, in
which there was a close passage between M82 and M81 (at a distance of 21
kpc), which started $\sim 500$ Myr ago (Brouillet et al.  1991), and
lasted for some 300 Myr (Yun 1999).  This independent dynamical estimate
of the last M81/M82 passage is remarkably close to the peak of the
cluster formation burst in M82 B.

We suggest therefore, that the last tidal encounter between M82 and M81
had a major impact on what was probably an otherwise normal, quiescent
disc galaxy.  It caused a concentrated burst of star formation, as
evidenced by the peak in the age distribution of the cluster sample in
M82 B.  Comparison of the cluster ages with the integrated light dating
in this region ($\sim 100-200$ Myr) suggests that field star formation
may have continued at a high rate after cluster formation had begun to
decline, but the uncertainties in the methods are large.  The enhanced
cluster formation decreased rapidly within a few hundred Myr of its
peak.  However, field star formation continued in M82 B, although
probably at a much lower rate, until $\sim 20$ Myr ago.  It has
evidently been suppressed during the last $\sim 10-15$ Myr, during which
the starburst in the core of M82 has been most active.  Evidence for
supernova remnants in the parts of M82 B nearest the starburst core
indicates that disc star formation during the last 50 Myr was more
active nearer the nucleus. 

At intermediate radii, between the active core and M82 B, Shen \& Lo
(1995) presented evidence for a change in the velocity dispersion
properties of the interstellar medium: they suggest that the much higher
velocity dispersion of the CO gas in the molecular spiral arms at $\sim
390$ pc from the core compared to the gas in the inner spiral arm at 125
pc indicates that the outer gas was disrupted by an earlier starburst. 
Although we cannot rule out the possibility that the starburst may in
fact have propagated inwards towards M82's core, this view is also
consistent with the dual-starburst hypothesis of Rieke et al.  (1993),
suggesting that the 30 Myr starburst occurred in the outer spiral arm
and the more recent $\sim 5$ Myr-old starburst in the inner arms, which
happened sufficiently recently for the interstellar medium to remain
relatively undisturbed. 

Satyapal et al.  (1997) presented a wealth of observational information
probing the central 500 pc, from which they concluded that there is a
significant age dispersion among the compact luminous infrared sources
(i.e., the embedded star clusters) within the central starburst, of
about 6 Myr.  Most interestingly, the age dispersion appears to be
correlated with projected distance from the core, leading these authors
to argue that the central starburst is possibly propagating outwards
towards the western edge of the starburst core, i.e., as a continuation
of the proposed propagation direction inferred from the overall
starburst geometry in M82. 

Independent H{\sc i} observations with MERLIN and the VLA (Pedlar \&
Wills, priv.  comm.) suggest a similar scenario: the compact sources
which correspond to young radio SNRs (whose progenitors have ages of
$\sim 10$ Myr), are much more extended throughout the central 750 pc of
the disc of M82 than the younger, $\sim 1$ Myr-old objects identified as
H{\sc ii} regions.  These latter sources are found to be mostly
concentrated towards the western edge of the galactic centre, in the
same sense as the star formation propagation direction proposed by
Satyapal et al.  (1997). 

Thus, the current starburst in the centre of the galaxy is either due to
large-scale propagating star formation throughout the disc of the galaxy
or possibly related to late infall of tidally disrupted debris from M82
itself, caused by dynamical feedback due to its gravitational potential
(O'Connell \& Mangano 1978, Yun et al.  1993).  It may, in fact, be a
combination of these two mechanisms, in the sense that the star
formation in the active core is actually propagating (cf.  Satyapal et
al.  1997), while -- as we suggested (de Grijs et al.  2001) -- the
overall starburst scenario is one in which the last tidal interaction
caused intense star formation in M82 B, of which the ejecta have
recently rained back onto the disc, causing the present-day starburst. 

Finally, a strong tidal interaction could easily produce an off-nuclear
starburst at a site like M82 B, although the M82 B burst could also have
been part of a larger scale event encompassing the centre of the galaxy
as well.  Given the high extinction and the dominance by much younger
concentrations of stars, it would not be easy to identify older clusters
in the starburst core, if they exist.  At the distance from the centre
of region B, one would expect M82's differential rotation (cf.  Shen \&
Lo 1995) to have caused the starburst area to disperse on these
time-scales.  The reason why the fossil starburst region has remained
relatively well constrained is likely found in the complex structure of
the disc.  It is well-known that the inner $\sim 1$ kpc of M82 is
dominated by a stellar bar (e.g., Wills et al.  2000) in solid-body
rotation.  From observations in other galaxies, it appears to be a
common feature that central bars are often surrounded by a ring-like
structure.  If this is also true for M82, it is reasonable to assume
that stars in the ring are trapped, and therefore cannot move very much
in radius because of dynamical resonance effects.  The phase mixing
around the ring might be slow enough for a specific part of the ring to
keep its identity over a sufficient time so as to appear like region B
(Gallagher, priv.  comm.): if the diffusion velocity around the ring is
sufficiently small, any specific region would remain self-constrained
for several rotation periods.  In addition, since the density in the
region is high, simple calculations imply that the area's self-gravity
is non-negligible compared to the rotational shear, therefore also
prohibiting a rapid dispersion of the fossil starburst region. 

\section{The Nature of the Young Compact Star Clusters in M82 B}

The evidence for decoupling between cluster and field star formation is
consistent with the view that young star cluster formation requires
special conditions, e.g., large-scale gas flows, in addition to the
presence of dense gas (cf.  Ashman \& Zepf 1992, Elmegreen \& Efremov
1997). 

Production of luminous, compact star clusters such as those described
above seems to be a hallmark of intense star formation episodes.  They
have been identified in several dozen galaxies, often involved in
interactions (e.g., Holtzman et al.  1992, Whitmore et al.  1993,
O'Connell et al.  1994, Conti et al.  1996, Watson et al.  1996, Carlson
et al.  1998, de Grijs et al.  2001).  Their sizes, luminosities, and --
in several cases -- masses are entirely consistent with what is expected
for young globular clusters (Meurer 1995, van den Bergh 1995, Ho \&
Filippenko 1996a,b, Ho 1997, Schweizer \& Seitzer 1998, de Grijs et al. 
2001). 

It is possible that a large fraction of the star formation in starbursts
takes place in the form of such concentrated clusters.  The discovery
that globular cluster formation, once thought to occur only during early
stages of galaxy evolution, continues today is one of the Hubble Space
Telescope's main contributions to astrophysics to date. 

Young compact star clusters are therefore important because of what they
can tell us about globular cluster formation and evolution (e.g.,
destruction mechanisms and efficiencies).  They are also important as
probes of the history of star formation, chemical evolution, initial
mass function, and other physical characteristics in starbursts.  This
is so because each cluster approximates a coeval, single-metallicity,
simple stellar population.  Such systems are the simplest to model, and
their ages and metallicities and, in some cases, initial mass functions
can be estimated from their integrated spectra. 

Using the individual age estimates obtained for the star clusters in M82
B, we can now apply the age-dependent mass-to-light ratio predicted for
a theoretical single burst stellar population to derive estimated masses
for this cluster sample.  Assuming a ``standard'' Salpeter (1955)
initial mass function, we find that the masses of the young clusters in
M82 B with $V \le 22.5$ mag are mostly in the range $10^4 - 10^6
M_\odot$, with a median of $10^5 M_\odot$. 

The high end of the M82 B cluster mass function overlaps with those
estimated by similar techniques for young compact star clusters in other
galaxies (e.g., Richer et al.  1993, Holtzman et al.  1996,
Tacconi-Garman et al.  1996, Watson et al.  1996, Carlson et al.  1998). 
Independent dynamical mass estimates are available only for a few of the
most luminous young star clusters, including M82 F, NGC 1569A and NGC
1705-1, and are approximately $10^6 M_\odot$ (Ho \& Filippenko 1996a,b,
Smith \& Gallagher 2001).  Because of the proximity of M82, we have been
able to probe the young cluster population in M82 B to fainter absolute
magnitudes, and thus lower masses, than has been possible before in
other galaxies.  Other young star cluster samples are biased toward high
masses by selection effects. 

The M82 B cluster masses are comparable to the masses of Galactic
globular clusters (e.g., Mandushev et al.  1991, Pryor \& Meylan 1993),
which are typically in the range $10^4 - 3 x 10^6 M_\odot$.  Some of the
clusters show evidence for asymmetries or subclustering, possibly an
indication of either mergers or extreme youth, implying that these
objects have not yet reached virial equilibrium.  The M82 B cluster
sizes are consistent with values found for the young compact star
cluster populations in the starburst core in M82 and other galaxies and
with the progenitors of globular clusters. 

If these clusters survive to ages of $\gtrsim 10$ Gyr, the M82 B
clusters will have properties similar to those of disc population
Galactic globulars.  However, the young globular cluster scenario would
be considerably strengthened if we could measure their masses directly. 
Recently, Smith \& Gallagher (2001) presented convincing evidence that
M82 F is likely dominated by a ``top-heavy'' present-day mass function,
truncated at $2-3 M_\odot$.  However, such claims, e.g., Rieke et al.'s
(1993) result for the main disc population in M82, have often turned out
to be erroneous in retrospect (see the review by Scalo 1998).  While
variations in the initial mass function do occur, they occur mostly on
the scales of individual star forming regions, while the large-scale
initial mass function remains largely constant (e.g., Kroupa 2001). 
Smith \& Gallagher (2001) therefore caution that M82 F, and also the
massive young star cluster NGC 1705-1, may indeed be an intermediate and
high-mass anomaly, possibly caused by mass segregation effects at birth,
in an otherwise ``normal'' larger-scale star forming complex.  Either
way, the implication of such a localised variation in the present-day
(or initial) mass function would clearly be that such objects will not
survive to become old Galactic globular cluster analogues, simply
because they lack long-lived low-mass stars.  It is therefore expected
that they will dissolve within a few Gyr of their formation. 

Thus, although all integrated observational properties of these young
star clusters point towards a Galactic globular cluster-like evolution,
only high-dispersion spectroscopy will allow us to assess their
present-day (or perhaps their initial) mass function shape, and hence
their survival chances to become old Galactic globular cluster
analogues.

\end{document}